\documentclass[12pt]{article}
\usepackage{epsfig,float}
\usepackage{hyperref}%

\date{}
\begin{document}

\title{Conformal Bulk Fields, Dark Energy and Brane Dynamics}

\author{Rui Neves\footnote{E-mail: \tt
rneves@ualg.pt}\hspace{0.2cm}
  and
Cenalo Vaz\footnote{E-mail: \tt cvaz@ualg.pt}\\
{\small \em Faculdade de Ci\^encias e Tecnologia,
Universidade do Algarve}\\
{\small \em Campus de Gambelas, 8000-117 Faro, Portugal}
}

\maketitle

\begin{abstract}
In the Randall-Sundrum scenario we analyze the dynamics of a
spherically symmetric 3-brane when the bulk is filled with 
matter fields. Considering a global conformal transformation whose
factor is the $Z_2$ symmetric warp we find a new set of exact dynamical 
solutions for which gravity is bound to the brane. The set corresponds
to a certain class of conformal bulk fields. We discuss the geometries
which describe the dynamics on the brane of polytropic dark energy. 
\end{abstract}

\section{Introduction}

In the Randall-Sundrum (RS) brane world scenario the visible
Universe is a 3-brane boundary of a $Z_2$ symmetric 5-dimensional
anti-de Sitter (AdS) orbifold \cite{RS1,RS2}. There are two basic settings. On one
hand the RS1 model \cite{RS1} with two branes and a compactified 
fifth dimension and on the other the RS2 model \cite{RS2} with a
single brane which may be associated with an infinite fifth dimension.  
With two branes the hierarchy problem is reformulated introducing an 
exponential warp in the fifth dimension. The gravitational
field is localized on the hidden positive tension brane and decays
towards the visible negative tension brane thus producing an
exponential hierarchy between the Planck and weak energy scales. In the RS2
model the same warping of the fifth dimension ensures that the
graviton is bound to the brane. 

In the RS models the
classical field dynamics is defined by 5-dimensional Einstein equations with
a negative bulk
cosmological constant $\Lambda_B$, Dirac delta sources representing
the branes and a stress-energy tensor describing other bulk field
modes \cite{RS1}-\cite{GW}. In the RS2 model a set of vaccum solutions is given by 

\begin{equation}
d{\tilde{s}_5^2}=
d{y^2}+{e^{-2|y|/l}}d{s_4^2}\;,
\end{equation}
where the 4-dimensional line element $d{s_4^2}$ is Ricci flat, $l$ is
the AdS radius given by $l=1/\sqrt{-{\Lambda_B}{\kappa_5^2}/6}$ with
${\kappa_5^2}=8\pi/{M_5^3}$ and $M_5$ the fundamental 5-dimensional 
Planck mass. The brane cosmological
constant is fine-tuned to be zero 
giving ${\Lambda_B}=-{\kappa_5^2}{\lambda^2}/6$ where $\lambda>0$
denotes the brane tension. Due to the periodicity and the $Z_2$
symmetry of the RS orbifold these solutions also hold for the RS1
model. Then the two branes have opposite tensions and $\lambda>0$ is
the tension of the hidden Planck brane.      

The low energy theory of gravity on the brane is 
4-dimensional general relativity and the cosmology may be 
Friedmann-Robertson-Walker \cite{RS1}-\cite{TM}. However, it should
be noted that in the
RS1 model this has only been achieved if a scalar field is introduced in the
bulk to stabilize the size of the fifth dimension
\cite{GW,WFGK,CGRT,TM}. The 
problem of the 
gravitational collapse of matter was also investigated in the RS scenario
\cite{CHR}-\cite{RC2}. It was found that a black string solution
first discussed in a different context by Myers and Perry \cite{MP} 
induced the Schwarzschild metric on the brane \cite{CHR}. However for
such a solution the
Kretschmann scalar diverged both at the AdS horizon and at the black string
singularity \cite{CHR}. The solution is thus expected to be unstable 
\cite{CHR,GL}. It may decay to a black
cylinder localized near the brane which is free from naked
singularities \cite{CHR}. So far this
continues to be a conjecture. Indeed, while exact solutions interpreted
as static black holes localized on a brane have been found for a
2-brane embedded in a 4-dimensional AdS space \cite{EHM}, a static
black hole localized on a 3-brane remains unknown. The difficulty lies
in the simultaneous localization of gravity and matter near the brane 
without the creation of naked singularities in the bulk
\cite{CHR}, \cite{KOP}-\cite{RC2}. This has inspired another conjecture
stating that $D+1$-dimensional black hole solutions localized on
a $D-1$-brane should correspond to quantum corrected $D$-dimensional
black holes on the brane \cite{EFK}. Related to the AdS/CFT
correspondence \cite{ADS/CFT} this connection 
provides an extra motivation to look for 5-dimensional collapse
solutions localized on a brane. There are also many braneworld
solutions, obained
from the effective 4-dimensional point of view by the application
of the covariant Gauss-Codazzi formulation \cite{BC,SMS}, that have so far not
been associated with exact 5-dimensional spacetimes \cite{COSp}-\cite{RC1}.

In this proceedings we report
on research about the dynamics of a 
spherically symmetric 3-brane when conformal fields are present in the bulk
\cite{RC2,RC3} (see also \cite{EO}). We
focus on 5-dimensional solutions which describe the dynamics of
polytropic dark energy on the brane \cite{RC3}. 

\section{Einstein Equations}

Let us start by introducing $(t,r,\theta,\phi,z)$ as coordinates in the 
5-dimensional bulk. The most general metric consistent with the 
$Z_2$ symmetry in $z$ and with 4-dimensional
spherical symmetry on the brane is given by  

\begin{equation}
d{\tilde{s}_5^2}={\Omega^2}\left(-{e^{2A}}d{t^2}
+{e^{2B}}d{r^2}+{R^2}d{\Omega_2^2}+d{z^2}\right)\;,
\end{equation}
where $\Omega=\Omega(t,r,z)$, $A=A(t,r,z)$, $B=B(t,r,z)$ and
$R=R(t,r,z)$ are $Z_2$ symmetric functions. $\Omega$ is the warp factor and
$R$ the physical radius of the 2-spheres. With a single brane the
classical dynamics is defined by

\begin{equation}
{\tilde{G}_\mu^\nu}=-{\kappa_5^2}\left[{\Lambda_B}{\delta_\mu^\nu}+
{\lambda\over{\sqrt{\tilde{g}_{55}}}}\delta
\left(z-{z_0}\right)\left(
{\delta_\mu^\nu}-{\delta_5^\nu}{\delta_\mu^5}\right)-
{\tilde{T}_\mu^\nu}\right]\;,
\label{5DEeq}
\end{equation}
where the bulk stress-energy tensor is conserved in the bulk

\begin{equation}
{\tilde{\nabla}_\mu}{\tilde{T}_\nu^\mu}=0\;.\label{5Dceq}
\end{equation}

For a general 5-dimensional metric $\tilde{g}_{\mu\nu}$ the Einstein
equations (\ref{5DEeq}) are extremely complex. To be able to solve
them we need simplifying assumptions about the field variables
involved in the problem. Let us first assume that under the 
conformal transformation ${\tilde{g}_{\mu\nu}}={\Omega^2}{g_{\mu\nu}}$
the bulk stress-energy tensor has conformal weight $s$,

\begin{equation} 
{{\tilde{T}}_\mu^\nu}={\Omega^{s+2}}{T_\mu^\nu}\;.
\end{equation}
Then Eq. (\ref{5DEeq}) may be re-written as 

\begin{eqnarray}
{G_\mu^\nu}&=&-6{\Omega^{-2}}\left({\nabla_\mu}\Omega\right){g^{\nu\rho}}
{\nabla_\rho}\Omega+
3{\Omega^{-1}}{g^{\nu\rho}}{\nabla_\rho}{\nabla_\mu}\Omega
-3{\Omega^{-1}}{\delta_\mu^\nu}{g^{\rho\sigma}}{\nabla_\rho}{\nabla_\sigma}
\Omega\nonumber\\
&&-{\kappa_5^2}
{\Omega^2}\left[{\Lambda_B}{\delta_\mu^\nu}+\lambda{\Omega^{-1}}
\delta(z-{z_0})\left(
{\delta_\mu^\nu}-{\delta_5^\nu}{\delta_\mu^5}\right)
-{\Omega^{s+2}}{T_\mu^\nu}\right]\;.
\label{t5DEeq}
\end{eqnarray}
Similarly under the conformal transformation the conservation equation becomes

\begin{equation}
{\nabla_\mu}{T_\nu^\mu}+{\Omega^{-1}}\left[(s+7){T_\nu^\mu}{\partial_\mu}
\Omega-{T_\mu^\mu}{\partial_\nu}\Omega\right]=0\label{t5Dceq}\;.
\end{equation}
If in addition it is assumed that
${{\tilde{T}}_\mu^\nu}={\Omega^{-2}}{T_\mu^\nu}$ then
Eq. (\ref{t5DEeq}) may be separated in the following way

\begin{equation}
{G_\mu^\nu}={\kappa_5^2}{T_\mu^\nu}\;,\label{r5DEeq}
\end{equation}
\begin{eqnarray}
&6{\Omega^{-2}}{\nabla_\mu}\Omega{\nabla_\rho}
\Omega{g^{\rho\nu}}-
3{\Omega^{-1}}{\nabla_\mu}{\nabla_\rho}\Omega{g^{\rho\nu}}+3{\Omega^{-1}}
{\nabla_\rho}{\nabla_\sigma}\Omega{g^{\rho\sigma}}{\delta_\mu^\nu}=\nonumber\\
&-{\kappa_5^2}
{\Omega^2}\left[{\Lambda_B}{\delta_\mu^\nu}+\lambda{\Omega^{-1}}
\delta(z-{z_0})\left(
{\delta_\mu^\nu}-{\delta_5^\nu}{\delta_\mu^5}\right)\right]\;.\label{5DEeqwf}
\end{eqnarray}
Because of the Bianchi identity we must also have

\begin{equation}
{\nabla_\mu}{T_\nu^\mu}=0\;,\label{r5Dceq}
\end{equation}
\begin{equation}
3{T_\nu^\mu}{\partial_\mu}\Omega-T {\partial_\nu}\Omega=0\;.
\label{5Dceqwf}
\end{equation}
Note that Eqs. (\ref{r5DEeq}) and (\ref{r5Dceq}) are 5-dimensional
Einstein equations with matter fields present in the bulk but without
a brane or bulk cosmological constant. They do not depend on the conformal
warp factor which is dynamically defined by Eqs. (\ref{5DEeqwf}) and 
(\ref{5Dceqwf}). The warp is then the only effect reflecting the
existence of the brane or of the bulk cosmological constant. We
stress that this is only possible for the special class of bulk fields
which have a stress-energy tensor with conformal weight $s=-4$.

Thought now partially decoupled the 5-dimensional Einstein equations
are still difficult to solve. Note for instance that the warp depends 
non-linearly on the  metric functions $A$, $B$ and $R$. So let us
further assume that $A=A(t,r)$, $B=B(t,r)$,
$R=R(t,r)$ and $\Omega=\Omega(z)$. Then we obtain 

\begin{equation}
{G_a^b}={\kappa_5^2}{T_a^b},\quad{\nabla_a}{T_b^a}=0\;,\label{4DEeq}
\end{equation}
\begin{equation}
{G_z^z}={\kappa_5^2}{T_z^z}\;,\label{5DEeqz}
\end{equation}
\begin{equation}
6{\Omega^{-2}}{{({\partial_z}\Omega)}^2} =
-{\kappa_5^2}{\Omega^2}{\Lambda_B}\;,
\end{equation}
\begin{equation}
3{\Omega^{-1}}{\partial_z^2}\Omega = -{\kappa_5^2}{\Omega^2}
\left[{\Lambda_B}+\lambda{\Omega^{-1}}\delta(z-{z_0})\right]\label{rswf}
\end{equation}
and \cite{KKOP,IR}
\begin{equation}
2{T_z^z}={T_c^c}\;,\label{eqst2}
\end{equation}
where the latin indices represent the coordinates $t,r,\theta$ and
$\phi$. 

The warp is now independent of $A$, $B$ and $R$. It may be
chosen to be the RS factor \cite{RS1,RS2,CHR}

\begin{equation}
\Omega={\Omega_{\mbox{\tiny RS}}}\equiv {l\over{|z-{z_0}|+{z_0}}}\;.\label{RSwf1}
\end{equation}
Naturally, other warp factors depending only on $z$ such as those
of thick branes \cite{CEHS} and non-fine-tuned branes \cite{KR} may 
also be considered (see \cite{KT}). On the other hand the highly coupled 
5-dimensional collapse dynamics has been 
reduced to 4-dimensional dynamics. Consider a diagonal stress-tensor,

\begin{equation}
{T_\mu^\nu}=diag\left(-\rho,{p_r},{p_T},{p_T},{p_z}\right)\;,\label{bmten}
\end{equation}
where $\rho$, $p_r$, $p_T$ and $p_z$ denote the bulk matter density and
pressures. Then Eq. (\ref{eqst2}) is
re-written as
\begin{equation}
\rho-{p_r}-2{p_T}+2{p_z}=0\;.\label{eqst3}
\end{equation}
The collapse of the conformal bulk matter is in general 
inhomogeneous
and defined by Eq. (\ref{4DEeq}). The matter dynamics  generates
a pressure $p_z$ along the fifth
dimension which must consistently be given by Eqs. (\ref{5DEeqz}) and
(\ref{eqst3}).     

\section{Polytropic Dark Energy}

To behave as polytropic dark energy the stress-energy tensor should be
of the form (\ref{bmten}) where the bulk matter density $\rho$ and
pressures $p_r$, $p_T$ and $p_z$ are given by

\begin{equation}
\rho={\rho_{\mbox{\tiny P}}},\;
{p_r}=-\eta{{\rho_{\mbox{\tiny P}}}^\alpha},\;{p_T}={p_r},\;{p_z}=-{1\over{2}}
\left({\rho_{\mbox{\tiny P}}}
+3\eta{{\rho_{\mbox{\tiny P}}}^\alpha}\right)\;.
\label{dmeqst}
\end{equation}
Above ${\rho_{\mbox{\tiny P}}}$ defines the polytropic dark energy
density and the parameters ($\alpha$, $\eta$) characterize
different polytropic phases. In what follows we restrict our attention
to the generalized Chaplygin phase characterized by $-1\leq\alpha<0$ 
\cite{RC2,KMP,BBS}. 

The polytropic energy density ${\rho_{\mbox{\tiny P}}}$ is obtained
by solving the conservation equations in Eq. (\ref{4DEeq}) which in
this case reads

\begin{equation}
\dot{{\rho_{\mbox{\tiny P}}}}+\left(\dot{B}+2{{\dot{R}}\over{R}}\right)
\left({\rho_{\mbox{\tiny P}}}-\eta{{\rho_{\mbox{\tiny P}}}^\alpha}\right)=0,\;
A'\left({\rho_{\mbox{\tiny P}}}-\eta{{\rho_{\mbox{\tiny P}}}^\alpha}\right)
-\eta\alpha{{\rho_{\mbox{\tiny P}}}^{\alpha-1}}{\rho_{\mbox{\tiny P}}}'=0\;.
\label{4Dceq}
\end{equation}
Taking an homogeneous density $\rho_{\mbox{\tiny P}} =
\rho_{\mbox{\tiny P}}(t)$, the metric function $A(t,r)$
may be safely set to zero. Then since the off-diagonal
Einstein equation ${G_t^r}=0$ has solution ${e^B}=R'/H$ with $H=H(r)$
an arbitrary function of $r$ we obtain

\begin{equation}
{\rho_{\mbox{\tiny P}}}={{\left(\eta+{a\over{S^{3-3\alpha}}}\right)}^
{1\over{1-\alpha}}}\;,\label{dmdena}
\end{equation}
where $a$ is an integration constant and $S=S(t)$ is the
Robertson-Walker scale factor of the brane world which is related to
the physical radius by $R=rS$. 

At small $S$ the Chaplygin dynamics is
dominated by the homogeneous dust phase with
${\rho_{\mbox{\tiny P}}}={a^{1/(1-\alpha)}}/{S^3}$. The Chaplygin gas
has an intermediate phase defined by the equation of
state ${p_{\mbox{\tiny P}}}=-\alpha{\rho_{\mbox{\tiny P}}}$ which
satisfies the dominant energy condition. For large $S$ the
dynamics is dominated by an effective cosmological constant term. An 
evolution of the Chaplygin equation of state may thus describe
the change in the dark energy behavior during the expansion of the 
Universe.

Next consider the diagonal Einstein equations in
Eq. (\ref{4DEeq}). Because ${p_r}={p_T}$ it must be ${G_r^r}=
{G_\theta^\theta}$. As a consequence 

\begin{equation}
{H^2}=1-k{r^2}\;,\label{HRWeq}
\end{equation}
where the constant $k$ is the Robertson-Walker curvature
parameter. Then substituting $R=rS$ in

\begin{equation}
{-G_t^t}+{G_r^r}+2{G_\theta^\theta}=-2{{\ddot{R}'}\over{R'}}-
4{{\ddot{R}}\over{R}}={\kappa_5^2}\left({\rho_{\mbox{\tiny P}}}-3\eta
{{\rho_{\mbox{\tiny P}}}^\alpha}\right)\label{4DEteq}
\end{equation}
we obtain

\begin{equation}
{\ddot{S}\over{S}}=-{{\kappa_5^2}\over{6}}\left({\rho_{\mbox{\tiny P}}}-3\eta{{\rho_{\mbox{\tiny
          P}}}^\alpha}\right)\;.\label{dmdyeq}
\end{equation}
On the other hand the radial equation ${G_r^r}={\kappa_5^2}{p_r}$ leads to

\begin{equation}
{\dot{S}^2}={{\kappa_5^2}\over{3}}{\rho_{\mbox{\tiny
        P}}}{S^2}-k\;.\label{dmdyeqa}
\end{equation}
Naturally, Eqs. (\ref{dmdyeq}) and (\ref{dmdyeqa}) are related by a
derivative. They are consistent when ${\rho_{\mbox{\tiny P}}}$
obeys the conservation Eq. (\ref{4Dceq}). Using Eqs. (\ref{dmdyeqa}), 
(\ref{dmdyeq}), (\ref{HRWeq}) and the expression for $p_z$ given in
Eq. (\ref{dmeqst}) we conclude that ${G_z^z}={\kappa_5^2}{p_z}$ is
an identity for all the parameters of the model. 

With $\Omega={\Omega_{\mbox{\tiny RS}}}$ given by Eq. (\ref{RSwf1}) we obtain the following 5-dimensional polytropic
solutions for which gravity is confined to the vicinity of the brane

\begin{equation}
d{\tilde{s}_5^2}={\Omega_{\mbox{\tiny RS}}^2}\left[-d{t^2}+{S^2}
\left({{d{r^2}}\over{1-k{r^2}}}+{r^2}d{\Omega_2^2}\right)+d{z^2}\right]\;.
\label{dmsol1}
\end{equation}
In Eq. (\ref{dmsol1}) $S$ satisfies Eq. (\ref{dmdyeqa}). Thought
obtained for the RS2 model these solutions also hold for the RS1 model
due to the periodicity and the $Z_2$ symmetry of the orbifold. In the 
RS1 model the two branes are then twin universes with opposite
tensions and an identical cosmological evolution. 

Let us now verify
that if a 4-dimensional observer confined to the
brane makes the same assumptions about the bulk degrees
of freedom then she deduces exactly the same dynamics
\cite{RC2}. Indeed, the non-zero components of the projected Weyl
tensor \cite{SMS} read

\begin{equation}
{\mathcal{E}_t^t}={{\kappa_5^2}\over{4}}\left({\rho_{\mbox{\tiny
        P}}}-\eta{\rho_{\mbox{\tiny P}}^\alpha}\right)
,\quad
{\mathcal{E}_r^r}={\mathcal{E}_\theta^\theta}={\mathcal{E}_\phi^\phi}=
-{{\mathcal{E}_t^t}\over{3}}\;.
\end{equation}
Then the effective 4-dimensional dynamics is given by

\begin{equation}
{G_t^t}=-{\kappa_5^2}{\rho_{\mbox{\tiny P}}},\quad
{G_r^r}={G_\theta^\theta}={G_\phi^\phi}=-{\kappa_5^2}\eta
{\rho_{\mbox{\tiny P}}^\alpha}\;.
\end{equation} 
The 4-dimensional observer
also sees gravity confined to the brane since she measures a negative
tidal acceleration \cite{RM} given by 

\begin{equation}
{a_T}={{{\kappa_5^2}{\Lambda_B}}\over{6}}\;.
\end{equation}
This implies that the geodesics just outside the brane converge towards
the brane and so for the 4-dimensional observer the conformal bulk
matter is effectively trapped inside the brane.

The effective 4-dimensional Chaplygin dynamics on the
brane may lead to the formation of a shell focusing singularity at
$S=0$ and of regular rebounce epochs at some $S\not=0$. This can be
analyzed \cite{RC1,DJCJ} with the following potential $V=V(S)$ defined by 

\begin{equation}
V(S)=S{\dot{S}^2}={{\kappa_5^2}\over{3}}{{\left(\eta{S^{3-3\alpha}}+
a\right)}^{1\over{1-\alpha}}}-k S\;.\label{dmdineq}
\end{equation}
If for all $S\geq 0$ it turns out that $V>0$ then a shell focusing
singularity forms at $S=0$. However, if an $S={S_*}>0$ exists such
that $V({S_*})=0$ then there is a regular rebounce point at $S={S_*}$. 

\begin{figure}[H]
\begin{center}
\includegraphics[width=.7\textwidth]{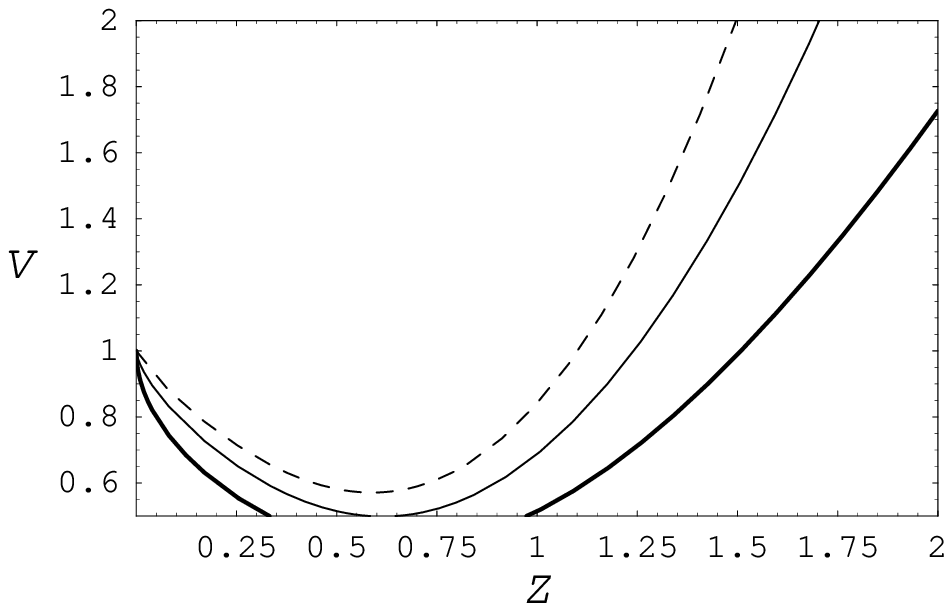}
\end{center}
\caption[]{Plots of $V$ for
$k>0$, $\eta>0$ and $a>0$. The dashed, thin and thick lines
correspond, respectively, to $\alpha$ equal to $-1/4,-1/2$ and $-1$}
\label{fig:l0ppp}
\end{figure}

According to recent experimental bounds \cite{BBS} the allowed range
of values for $\eta$ and $a$ are $\eta>0$ and $a>0$ or $\eta<0$ and
$a<0$. For $k=0$ the potential is positive for all $-1\leq\alpha<0$ if
$\eta>0$ and $a>0$. If $\eta<0$ and $a<0$ then this only happens for
the set $\alpha=-p/q$, $q>p$ with $q$ and $p$
, respectively, even and odd integers. In any of these cases there are only
singular solutions without rebouncing epochs. The Chaplygin
shells may either expand continously to infinity or collapse to the
singularity at $S=0$ where

\begin{equation}
V(0)={{\kappa_5^2}\over{3}}{{a}^{1\over{1-\alpha}}}>0\;.\label{V0}
\end{equation} 
It is for $k\not=0$ that 
new dynamics appears. With $S={Z^{1\over{1-\alpha}}}$ we find

\begin{equation}
V=V(Z)={{\kappa_5^2}\over{3}}{{\left(\eta {Z^3}+
a\right)}^{1\over{1-\alpha}}}-k{Z^{1\over{1-\alpha}}}\;.
\end{equation}
Consider $k>0$, $\eta>0$ and $a>0$ (see Fig.~\ref{fig:l0ppp}). The 
condition $V\geq 0$
is equivalent to ${\mathcal{V}}={\mathcal{V}}(Z)\geq 0$ where

\begin{equation}
{\mathcal{V}}={{\left({{\kappa_5^2}\over{3}}\right)}^{1-\alpha}}
\left(\eta{Z^3}+a\right)-{k^{1-\alpha}}Z\;.
\end{equation}
Then there are at most two regular rebounce epochs in the allowed
dynamical phase space. Since we have 

\begin{equation}
{\mathcal{V}}(0)={{\left({{\kappa_5^2}\over{3}}\right)}^{1-\alpha}}a>0,\;
{\mathcal{V}}'=3{{\left({{\kappa_5^2}\over{3}}\right)}^{1-\alpha}}\eta{Z^2}-{k^{1-\alpha}}
\end{equation}
and
\begin{equation}
{\mathcal{V}}''=6{{\left({{\kappa_5^2}\over{3}}\right)}^{1-\alpha}}\eta Z\geq 0
\end{equation}
this is determined by the
sign of $\mathcal{V}$ at its minimum
${{\mathcal{V}}_m}={\mathcal{V}}({Z_m})$ where
${Z_m}=\sqrt{
{{(3k/{\kappa_5^2})}^{1-\alpha}}/3\eta}$. 

If ${{\mathcal{V}}_m}>0$ then
there are no regular rebounce points and the
collapsing shells may fall from infinity to the singularity at
$S=0$ where $V(0)>0$ is given in Eq. (\ref{V0}). For
${{\mathcal{V}}_m}=0$ we have just one regular fixed point $S={S_*}$ which
divides the phase space into two disconnected regions, a bounded
region with the singularity at $S=0$, $0\leq S<{S_*}$, and an
infinitely extended region, $S>{S_*}$,
where the shells expand with ever increasing speed to infinity. In
this region the solutions are regular. If
${{\mathcal{V}}_m}<0$ then we have two regular rebounce epochs $S={S_-}$
and $S={S_+}$ such that ${S_-}<{S_+}$. 
For $0\leq S\leq{S_-}$ a shell may expand
to a maximum radius $r{S_-}$ and then rebounce to collapse towards the
singularity at $S=0$. For $S\geq {S_+}$ the collapsing shells shrink
to the minimum scale $S_+$ and then rebounce into accelerated
expansion to infinity. For $\eta<0$ and
$a<0$ we find the same type of dynamics but now only
for the special values $\alpha=-p/q$, $q>p$ with $q$ and $p$
, respectively, even and odd integers.

\begin{figure}[H]
\begin{center}
\includegraphics[width=.7\textwidth]{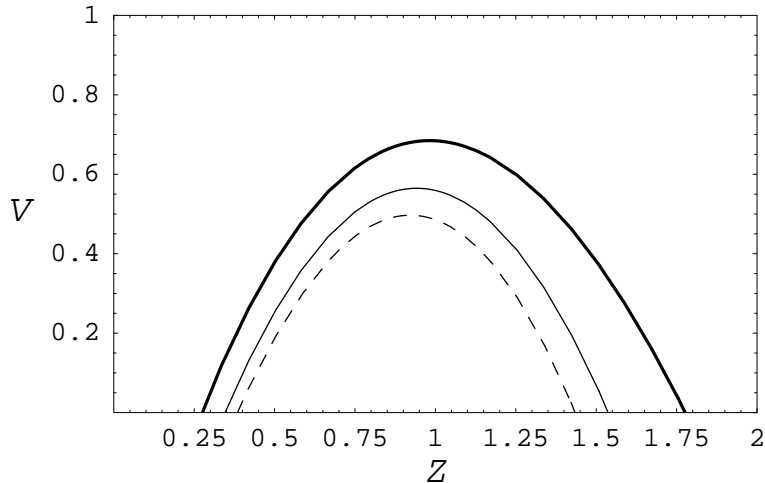}
\end{center}
\caption[]{Plots of $V$ for $k<0$, $\eta<0$ and $a<0$. The dashed, thin and thick lines
correspond, respectively, to $\alpha$ equal to $-2/7,-2/5$ and $-2/3$}
\label{fig:l0nnn}
\end{figure}

If $k<0$ then for $\eta>0$ and
$a>0$ the potential is always positive and so there are only singular
solutions without rebouncing points. For $\eta<0$ and $a<0$ (see
Fig.~\ref{fig:l0nnn}) we must
consider $\alpha=-p/q$, $q>p$
with $q$ and $p$, respectively, odd and even integers to find
solutions with rebounce epochs. The condition $V\geq 0$ is still
equivalent to ${\mathcal{V}}\geq 0$ but now

\begin{equation}
{\mathcal{V}}=-{{\left({{\kappa_5^2}\over{3}}\right)}^{1-\alpha}}
\left(|\eta|{Z^3}+|a|\right)+{{|k|}^{1-\alpha}}Z.
\end{equation}
Because we have 

\begin{equation}
{\mathcal{V}}(0)=-{{\left({{\kappa_5^2}\over{3}}\right)}^{1-\alpha}}|a|<0,\;
{\mathcal{V}}'=-3{{\left({{\kappa_5^2}\over{3}}\right)}^{1-\alpha}}|\eta|{Z^2}
+{{|k|}^{1-\alpha}}
\end{equation}
and

\begin{equation} 
{\mathcal{V}}''=-6{{\left({{\kappa_5^2}\over{3}}\right)}^{1-\alpha}}|\eta| Z\leq 0
\end{equation}
the sign of
$\mathcal{V}$ at its maximum
${{\mathcal{V}}_M}={\mathcal{V}}({Z_M})$ where
${Z_M}=\sqrt{
{{(3|k|/{\kappa_5^2})}^{1-\alpha}}/3|\eta|}$
shows that the only
possibilities are the existence of one or two regular
rebounce epochs.
In the former case
the classical brane stays forever in the fixed point and in the latter it
oscilates back and forth between the two rebounce points.

\section{Conclusions}

In this work we have presented new exact 5-dimensional solutions for
which gravity is localized in the vicinity of the brane and the
dynamics of the bulk fields on the brane is that of an homogeneous
Chaplygin gas, a possible candidate for the missing dark energy which
controls the expansion of the visible Universe. The bulk fields
were seen to belong to
a special conformal class which has a stress-energy tensor with
conformal weight $-4$. We have seen that the 5-dimensional solutions are
valid for the RS1 and the RS2 models and noted that an observer confined to the brane is led to the same
localized braneworld dynamics when using an identical
description of the field variables. We have analyzed the dynamical phase space
describing the evolution of the Chaplygin shells discussing conditions
for the formation of shell focusing singularities and of regular
rebounce epochs. However,
althought gravity is bound to the brane the conformal bulk
fields are not localized near the brane. Indeed, the density and
pressures increase with $z$ due to the scale factor ${\Omega^{-2}}(z)$
and diverge at the AdS horizon. This is not a problem in the RS1 model
because the space is cut before the AdS horizon is reached. In the RS2 model 
a solution requires the simultaneous localization of bulk matter and
gravity near the brane, an open problem for future research.    
\vspace{0.75cm}

\centerline{\bf Acknowledgements}
\vspace{0.25cm}

We are grateful for financial support from 
{\it Funda\c {c}\~ao para a Ci\^encia e a Tecnologia} (FCT) and {\it Fundo Social Europeu} (FSE) under the contracts
SFRH/BPD\-/7182/2001 and POCTI/32694/FIS/2000
({\it III Quadro Comunit\'ario de Apoio}) as well as from {\it Centro 
Multidisciplinar de Astrof\'{\i}sica} (CENTRA).


\begin{thebibliography}{30.}

\bibitem{RS1} L. Randall, R. Sundrum: Phys. Rev. Lett. \textbf {83}, 
3370 (1999)

\bibitem{RS2} L. Randall, R. Sundrum: Phys. Rev. Lett. \textbf
  {83}, 4690 (1999)

\bibitem{GW}
W. D. Goldberger, M. B. Wise: Phys. Rev. D. \textbf {60}, 107505 (1999)
Phys. Rev. Lett. \textbf {83}, 4922 (1999) Phys. Lett. B \textbf {475}, 275 (2000)

\bibitem{NK}
N. Kaloper: Phys. Rev. D \textbf {60}, 123506 (1999)\\
T. Nihei: Phys. Lett. B \textbf {465}, 81 (1999)\\
C. Cs\'aki, M. Graesser, C. Kolda, J. Terning: Phys. Lett. B
\textbf {462}, 34 (1999)\\
J. M. Cline, C. Grojean, G. Servant: Phys. Rev. Lett. \textbf {83},
4245 (1999)

\bibitem{KKOP}
P. Kanti, I. I. Kogan, K. A. Olive, M. Pospelov: Phys. Lett. B
\textbf {468}, 31 (1999) Phys. Rev. D \textbf {61}, 106004 (2000)

\bibitem{WFGK}
O. DeWolf, D. Z. Freedman, S. S. Gubser, A. Karch: Phys. Rev. D
\textbf {62}, 046008 (2000)

\bibitem{GT}
J. Garriga, T. Tanaka: Phys. Rev. Lett. \textbf {84}, 2778 (2000)

\bibitem{GKR}
S. Giddings, E. Katz, L. Randall: J. High Energy Phys. \textbf {03}, 023 (2000)

\bibitem{CGRT}
C. Cs\'aki, M. Graesser, L. Randall, J. Terning: Phys. Rev. D
\textbf {62}, 045015 (2000)

\bibitem{TM}
T. Tanaka, X. Montes: Nucl. Phys. \textbf {B582}, 259 (2000)

\bibitem{CHR}
A. Chamblin, S. W. Hawking, H. S. Reall: Phys. Rev. D \textbf {61}, 065007
(2000)

\bibitem{EHM}
R. Emparan, G. T. Horowitz, R. C. Myers: J. High Energy
Phys. \textbf {0001}, 007 (2000) J. High Energy Phys. \textbf {0001}, 021 (2000)

\bibitem{KOP}
P. Kanti, K. A. Olive, M. Pospelov: Phys. Lett. B \textbf {481}, 386
(2000)\\
T. Shiromizu, M. Shibata: Phys. Rev. D \textbf {62}, 127502 (2000)\\
H. L\"u, C. N. Pope: Nucl. Phys. \textbf {B598}, 492 (2001)\\
A. Chamblin, H. R. Reall, H. a. Shinkai, T. Shiromizu: Phys. Rev. D
\textbf {63}, 064015 (2001)\\
M. S. Modgil, S. Panda, G. Sengupta: Mod. Phys. Lett. A \textbf {17}
1479 (2002)

\bibitem{KT}
P. Kanti, K. Tamvakis: Phys. Rev. D \textbf {65}, 084010 (2002)

\bibitem{EFK}
 R. Emparan, A. Fabbri, N. Kaloper: J. High Energy Phys. \textbf
 {0208}, 043 (2002)\\
R. Emparan, J. Garcia-Bellido, N. Kaloper: J. High Energy
 Phys. \textbf {0301}, 079 (2003)

\bibitem{RC2}
R. Neves, C. Vaz: Phys. Rev. D \textbf {68}, 024007 (2003)

\bibitem{MP}
R. C. Myers, M. J. Perry: Ann. Phys. \textbf {172}, 304 (1986)

\bibitem{GL}
R. Gregory, R. Laflamme: Phys. Rev. Lett. \textbf {70}, 2837 (1993)
Nucl. Phys. \textbf {B428}, 399 (1994)

\bibitem{ADS/CFT}
J. Maldacena: Adv. Theor. Math. Phys. \textbf {2}, 231 (1998)\\
S. S. Gubser, I. R. Klebanov, A. M. Polyakov:
Phys. Lett. B \textbf {428}, 105 (1998)\\
E. Witten: Adv. Theor. Math. Phys. \textbf {2}, 253 (1998)

\bibitem{BC}
B. Carter: Phys. Rev. D \textbf {48}, 4835 (1993)\\
R. Capovilla, J. Guven: Phys. Rev. D \textbf {51}, 6736 (1995);
Phys. Rev. D \textbf {52}, 1072 (1995)

\bibitem{SMS}
T. Shiromizu, K. I. Maeda, M. Sasaki: Phys. Rev. D \textbf {62}, 024012
(2000)\\
M. Sasaki, T. Shiromizu, K. I. Maeda: Phys. Rev. D \textbf {62},
024008 (2000)

\bibitem{COSp}
J. Garriga, M. Sasaki: Phys. Rev. D \textbf {62}, 043523 (2000)\\
R. Maartens, D. Wands, B. A. Bassett , I. P. C. Heard: Phys. Rev. D
\textbf {62}, 041301 (2000)\\
H. Kodama, A. Ishibashi, O. Seto: Phys. Rev. D {62}, 064022
(2000)\\
D. Langlois: Phys. Rev. D \textbf {62}, 126012 (2000)\\
C. van de Bruck, M. Dorca, R. H. Brandenberger, A. Lukas: Phys. Rev.
D \textbf {62}, 123515 (2000)\\
K. Koyama, J. Soda: Phys. Rev. D \textbf {62}, 123502 (2000)\\
D. Langlois, R. Maartens, M. Sasaki, D. Wands: Phys. Rev. D \textbf
{63}, 084009 (2001)

\bibitem{DMPR}
N. Dadhich, R. Maartens, P. Papadopoulos, V. Rezania:
    Phys. Lett. B \textbf {487}, 1 (2000)\\
N. Dadhich, S. G. Ghosh: Phys. Lett. B \textbf {518}, 1 (2001)\\
C. Germani, R. Maartens: Phys. Rev. D \textbf {64}, 124010 (2001)\\
M. Bruni, C. Germani, R. Maartens: Phys. Rev. Lett. \textbf {87},
231302 (2001)\\
M. Govender, N. Dadhich: Phys. Lett. B \textbf {538}, 233 (2002)\\
R. Casadio, A. Fabbri, L. Mazzacurati: Phys. Rev. D \textbf {65},
084040 (2002)

\bibitem{RM}
R. Maartens: Phys. Rev. D \textbf {62}, 084023 (2000)

\bibitem{RC1}
R. Neves, C. Vaz: Phys. Rev. D \textbf {66}, 124002 (2002)

\bibitem{RC3}
R. Neves, C. Vaz: Phys. Lett. B \textbf {568}, 153 (2003)

\bibitem{EO}
E. Elizalde, S. D. Odintsov, S. Nojiri, S. Ogushi: Phys. Rev. D
\textbf {67}, 063515 (2003)\\
S. Nojiri, S. D. Odintsov: JCAP \textbf {06}, 004 (2003)

\bibitem{IR}
I. Z. Rothstein: Phys. Rev. D \textbf {64}, 084024 (2001)

\bibitem{CEHS}
C. Cs\'aki, J. Ehrlich, T. J. Hollowood, Y. Shirman: Nucl. Phys.
\textbf {B581}, 309 (2000)

\bibitem{KR}
A. Karch, L. Randall: J. High Energy Phys. \textbf {0105}, 008 (2001)

\bibitem{KMP}
A. Kamenshchik, U. Moschella, V. Pasquier: Phys. Lett. B \textbf {511},
265 (2001)\\
N. Bili\'c, G. B. Tupper, R. D. Viollier: Phys. Lett. B \textbf {535},
17 (2002)

\bibitem{BBS}
M. C. Bento, O. Bertolami, S. S. Sen: Phys. Rev. D \textbf {66},
043507 (2002) Phys. Rev. D \textbf {67}, 063003 (2003)

\bibitem{DJCJ}
S. S. Deshingkar, S. Jhingan, A. Chamorro, P. S. Joshi: Phys.
Rev. D \textbf {63}, 124005 (2001)

\end{thebibliography}
\end{document}